\documentclass[a4paper]{jpconf}
\usepackage{graphicx}
\begin{document}
\title{GPU-accelerated track reconstruction in the ALICE High Level Trigger}

\author{David Rohr$^{1,2}$, Sergey Gorbunov$^1$, Volker Lindenstruth$^{1,3}$ for the ALICE Collaboration}

\address{$^1$ Frankfurt Institute for Advanced Studies, Ruth-Moufang-Str. 1, 60438 Frankfurt, Germany\\
         $^2$ CERN, 385 Route de Meyrin, 1217 Meyrin, Switzerland\\
         $^3$ Institut f\"ur Informatik, Johann Wolfgang Goethe-Universit\"at Frankfurt, Robert-Mayer-Str. 11-15, 60629 Frankfurt, Germany\\}

\ead{drohr@compeng.uni-frankfurt.de}

\begin{abstract}
\looseness=-1
ALICE (A Large Heavy Ion Experiment) is one of the four major experiments at the Large Hadron Collider (LHC) at CERN.
The High Level Trigger (HLT) is an online compute farm which reconstructs events measured by the ALICE detector in real-time.
The most compute-intensive part is the reconstruction of particle trajectories called tracking and the most important detector for tracking is the Time Projection Chamber (TPC).
The HLT uses a GPU-accelerated algorithm for TPC tracking that is based on the Cellular Automaton principle and on the Kalman filter.
The GPU tracking has been running in 24/7 operation since 2012 in LHC Run 1 and 2.
In order to better leverage the potential of the GPUs, and speed up the overall HLT reconstruction, we plan to bring more reconstruction steps (e.g. the tracking for other detectors) onto the GPUs.
There are several tasks running so far on the CPU that could benefit from cooperation with the tracking, which is hardly feasible at the moment due to the delay of the PCI Express transfers.
Moving more steps onto the GPU, and processing them on the GPU at once, will reduce PCI Express transfers and free up CPU resources.
On top of that, modern GPUs and GPU programming APIs provide new features which are not yet exploited by the TPC tracking.
We present our new developments for GPU reconstruction, both with a focus on the online reconstruction on GPU for the online offline computing upgrade in ALICE during LHC Run~3, and also taking into account how the current HLT in Run 2 can profit from these improvements.
\end{abstract}

\section{Introduction}
\begin{figure}[t]
\begin{minipage}{16.5pc}
\includegraphics[width=16.5pc]{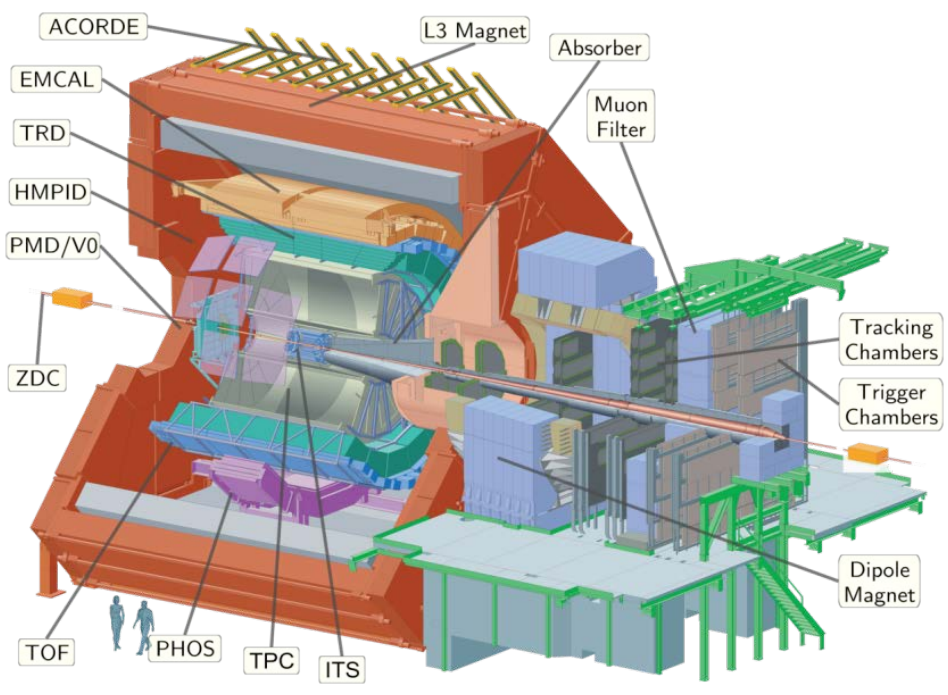}
\caption{The ALICE apparatus and its various detector components.}
\label{fig:alice}
\end{minipage}\hfill
\begin{minipage}{17.5pc}
\includegraphics[width=17.5pc]{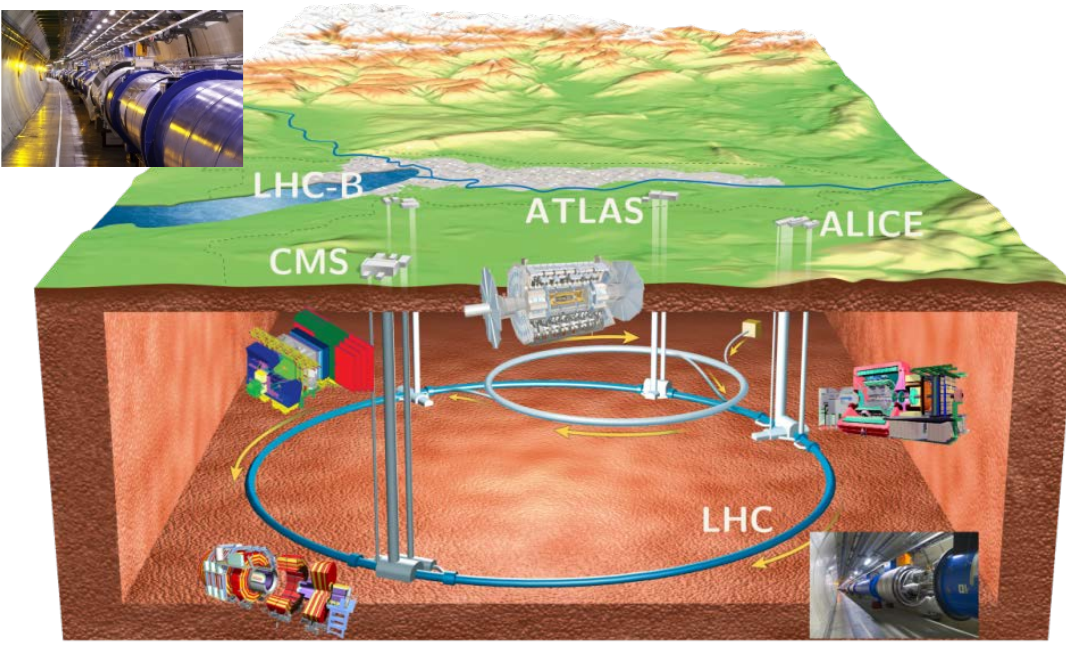}
\caption{The Large Hadron Collider beneath Geneva and its four major experiments.}
\label{fig:lhc}
\end{minipage}\hspace{2pc}
\end{figure}

\looseness=-1
ALICE (A Large Ion Collider Experiment~\cite{alice}, see Figure~\ref{fig:alice}) is one of four large-scale experiments at the LHC (Large Hadron Collider, see Figure~\ref{fig:lhc}) at CERN in Geneva.
It is designed primarily to record heavy ion collisions.
Compared to proton-proton collisions, which are the focus of the other LHC experiments, heavy ion collisions produce many more particles but occur at a lower rate.
The ALICE High Level Trigger (HLT) is an online compute farm that processes the data recorded by the ALICE detectors in real time.
Reconstructing the particle trajectories (tracking) is the most compute intense task of event processing.
ALICE's primary detector for tracking is the TPC (Time Projection Chamber).
This drift chamber records up to 159 hits per trajectory traversing its volume, which are significantly more data points compared to the silicon detectors used in the ALICE Inner Tracking System (ITS) and by the other LHC experiments.
Consequently, Pb-Pb events recorded by ALICE are larger by more than one order of magnitude compared to pp events.
In contrast, ALICE records heavy ion collisions at a much lower rate of up to~$1$\,kHz.
This has an impact on the tracking algorithm.
The ALICE HLT employs GPU-accelerated tracking which enables it to reconstruct all events in real time~\cite{gpu}.

\looseness=-1
The HLT GPU tracking was originally developed for NVIDIA GPUs in the HLT farm for LHC Run~1.
Utilization of GPUs accelerated the track reconstruction by around a factor of~$10$ compared to a CPU solution running on four CPU cores~\cite{gpu2}.
We note that the CPU version does not use explicit vectorization, because attempts to do so yielded only a speedup of up to 1.25 for central Pb-Pb collisions due to varying track length~\cite{kretz}.
This aspect also significantly reduces GPU utilization~\cite{tns}.
Overall, the GPU approach could essentially halve the number of required HLT compute nodes compared to a CPU-only solution.
Consequently, power consumption goes down significantly, but we did not conduct precise power efficiency measurements yet.
For the HLT Run~2 farm the GPU tracking algorithm was ported to support OpenCL, CUDA, and OpenMP for CPUs to become vendor independent.
The GPU tracker is currently running on AMD S9000 GPU since 2015~\cite{chep15}.
New CPU and GPU models have improved their performance similarly, such that also for the Run~2 farm the GPU tracking halves the number of compute nodes.
The total cost saving of the GPU tracking compared to a CPU-only cluster has accumulated to more than~$1.5$ million Swiss francs.
In order to better understand the new optimizations we are working on, we give an overview of the tracking algorithm.

\section{Tracking algorithm}

\begin{figure}[b]
\includegraphics[width=1.0\textwidth]{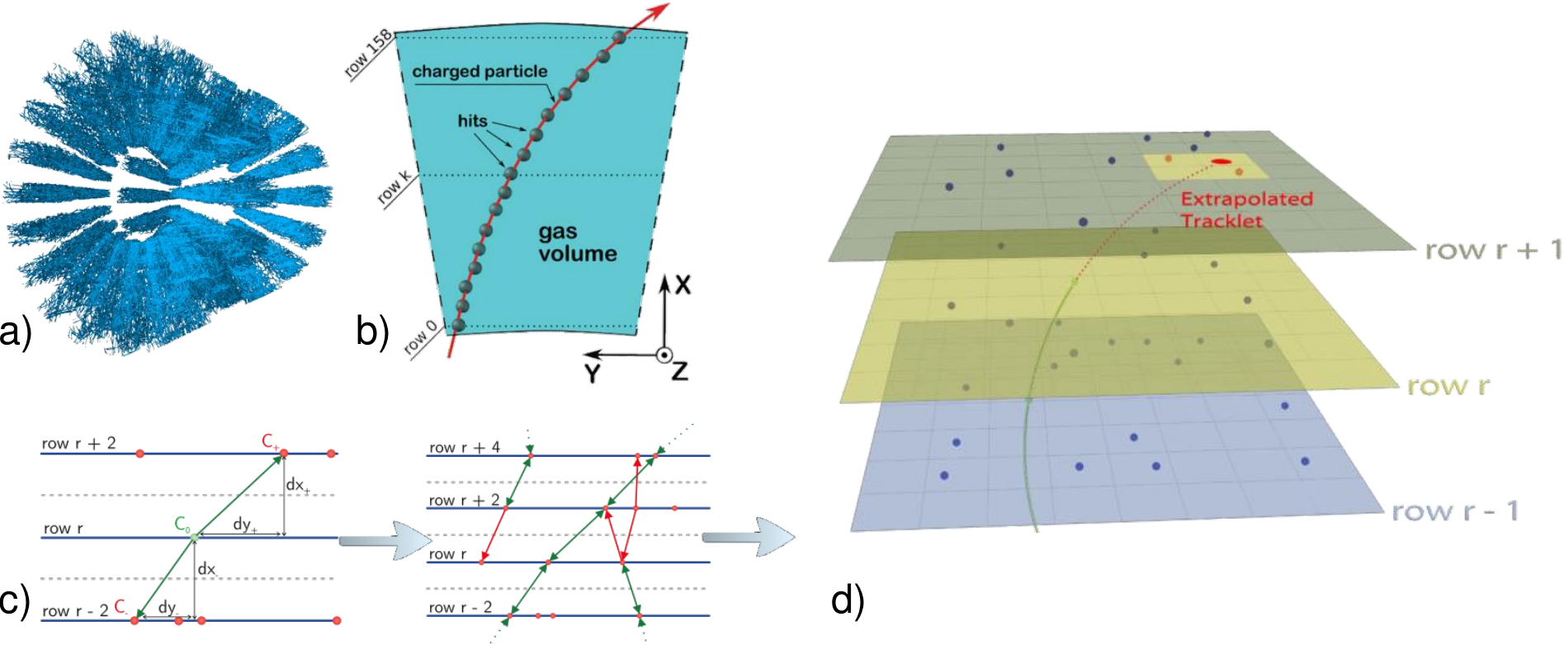}
\begin{minipage}[b]{1.0\textwidth}\caption{Overview of the ALICE HLT TPC GPU tracking algorithm: a) Separation of the TPC into sectors. b) Cross section of a TPC sector. c) Cellular automaton heuristic to construct initial track seeds. d) Track following and fitting using the Kalman filter in radial TPC pad-row direction, adding the closest cluster in each two-dimensional (pad and time) row.}
\label{fig:d}
\end{minipage}
\end{figure}

The tracking is organized in two phases:
the first phase searches only for track segments inside TPC sectors.
The TPC cylinder is divided into two halves which are further subdivided into~$18$ trapezoidal sectors each as shown in Figure~\ref{fig:d}~a).
Within the sector, primary trajectories are assumed to have at least a part of the trajectory oriented in~$x$-direction (see Figure~\ref{fig:d}~b)).
The second phase merges track segments, first within sectors and then at sector borders, and finally performs the full track fit.
Both phases have many sub-steps.
We list only the important and compute-intense steps (Table~\ref{tab1} gives an overview).
\begin{enumerate}
 \item \textbf{Seeding (Phase I):}
   Finds short track candidates of usually~$3$ to about~$10$ clusters using a heuristics in a cellular automaton.
   It finds only seeds oriented along the~$x$-axis, but due to the large amount of hits per track in the TPC, this is no limitation.
   In addition, there are so many seeds that it is not necessary to follow multiple track hypotheses per seed, which reduces the combinatorics significantly.
   Figure~\ref{fig:d}~c) illustrates the process.
 \item \looseness=-1 \textbf{Track following (Phase I):}
   Fits parameters to the track candidate and extrapolates the track through the TPC sector volume to find all hits of the track segment.
   The track parameters are regularly refined by fitting the new hits added to the track.
   A simplified Kalman filter is used which assumes uncorelated errors for coordinates and a constant magnetic field to speed up this process.
   This step only finds the tracks, the final fit is done at a later stage.
   In case multiple seeds were found and processed for the same track segment, the best instance is selected and others are removed.
   Figure~\ref{fig:d}~d) illustrates the process.
 \item \textbf{Track merging (Phase II):}
   This step creates the final tracks by merging track segments.
   First, the tracker merges track segments inside the sector for the case that there is a gap which the track following did not cover.
   Finally, track parameters at the border of the sectors are compared and matching tracks are merged.
 \item \textbf{Track fit:}
   In the end, the full track is refit using the full Kalman filter.
   A polynomial approximation of the magnetic field avoids the slow memory access to a field map.
\end{enumerate}

\begin{table}
\centering
\caption{\label{tab1}Steps of the ALICE HLT TPC GPU tracking.
Tracking runs in two phases.
Both phases have several steps.
The table shows only steps with significant contribution to CPU time.
The time is measured with the CPU-only version single-threaded.
GPU processing of step~$4$ is currently not used.}
\begin{tabular}{lllllrl}
\br
\# & Phase & Task & Method & Locality & Time & Device \\
\mr
1 & I  & Seeding & Cellular Automaton & Very local & $30$\,$\%$ & CPU \& GPU \\
2 &    & Track following & Simple Kalman filter & Sector-local & $60$\,$\%$ & CPU \& GPU \\
\mr
3 & II & Track Merging & Matching Covariance & Global & $2$\,$\%$ & CPU \\
4   &  & Final Fit & Kalman filter & Global & $8$\,$\%$ & CPU (or GPU) \\
\br
\end{tabular}
\end{table}

The first phase has been completely ported to GPUs and is running in 24/7 operation in the HLT since the end of 2012.
From the second phase, only the compute intense track fit part has been ported as proof of concept, which is well suited employing a full compute-intense Kalman filter.
Even though the GPU speeds up this step by an order of magnitude, it yields no significant gain in reality because the additional data transfer takes much longer than the actual processing.
If the remaining steps were to be ported on the GPU, the GPU track fit could speed up the processing (compare Figure~\ref{fig:b} in the next section).

A great advantage of the tracking approach is the fact that only one track hypothesis at max is followed per hit.
Consequently, we could show that the tracking time depends almost completely linearly on the input data size~\cite{gpu}.
This is very different to tracking algorithms for other detector types and the other LHC experiments, which face super-linear dependencies.
However, with the data sizes of ALICE Pb-Pb events, a quadratic dependency on the input data size would be an immediate show-stopper for online track reconstruction.

\section{GPU tracking optimizations}

The current HLT farm provides sufficient GPU resources to perform the online tracking for any scenario during run~$2$.
We have tested the farm with data replay of pp and Pb-Pb TPC input at the maximum possible link speed from the detector~\cite{framework}.
Therefore, no further optimization particularly for the Run~2 hardware is needed.
Instead, we focus on improving the tracking in the face of the upcoming ALICE online offline computing upgrade for LHC Run~3.

\begin{figure}[t]
\includegraphics[width=1.0\textwidth]{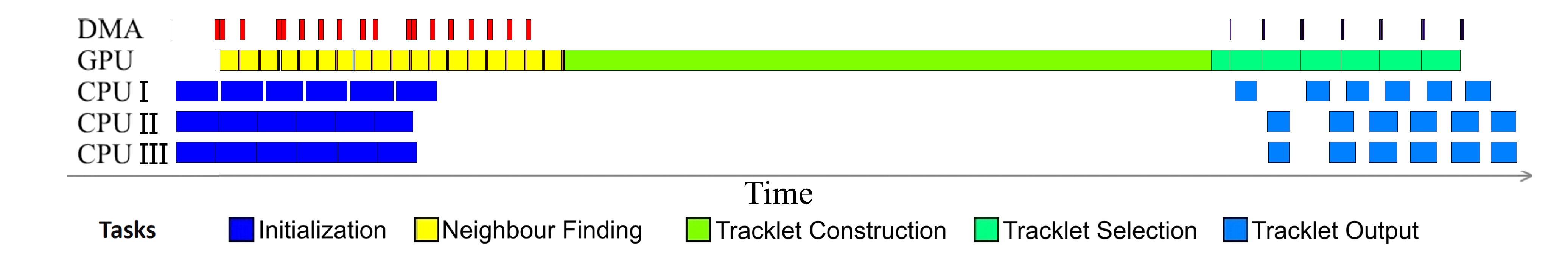}
\begin{minipage}[b]{1.0\textwidth}\caption{Visualization of the parallel processing of phase I on CPU and GPU of the ALICE track reconstruction during Run~1 and beginning of Run~2.
Times measured on NVIDIA GTX 480 GPU.
Three CPU cores are used for the preprocessing and postprocessing steps in a round robin fashion since one core is too slow.}
\label{fig:async}
\end{minipage}
\end{figure}

\looseness=-1
In order to achieve good GPU utilization, the tracker overlaps computation on the GPU, DMA transfer to and from the GPU, as well as preprocessing and postprocessing on the host.
Figure~\ref{fig:async} shows the time and overlap for all tasks on an NVIDIA GTX 480 GPU.
The old GPUs could not run two different kernels concurrently, therefore two CUDA streams or OpenCL command queues respectively are used for computation and for DMA transfer.
Every box in the plot corresponds to a kernel execution.
It starts with the initialization on the host with one box per TPC sector, followed by the neighbor finding (part of the cellular automaton) on the GPU, also with one kernel per sector.
Processing of the sectors is arranged in a pipeline, such that after the preprocessing on the host and the transfer of the first sector to the GPU, the graphics card is constantly loaded.
However, this only shows that the GPU is loaded during~$100$\,$\%$ of the time, but it does not guarantee that each kernel individually uses the GPU to the full extent.
The cellular automaton runs one GPU thread per hit in the TPC, therefore one sector offers sufficient parallelism to fully load the GPU.
This is not the case during the tracklet construction, which performs the track following inside a sector.
With one thread per track candidate, there are insufficient tracks per sector to use the GPU efficiently.
Therefore, only one instance of the kernel runs (one box) processing all sectors.
Processing the tracklet selection is similar to the cellular automaton, however, few sectors are combined into one kernel to improve GPU utilization.

This approach was optimal for older GPU models but it is no longer sufficient to utilize modern graphics cards at maximum.
A single sector is lacking parallelism for the cellular automaton, and a single event cannot load the GPU at maximum during track following.
Parallelization is insufficient even for the most central Pb-Pb events.
Therefore, it is desirable to execute multiple kernels concurrently, which is supported by all recent GPUs.
We can realize this in two ways, which are described in the following.

\subsection{Parallelization inside an event}

\begin{figure}[t]
\begin{center}\includegraphics[width=0.7\textwidth]{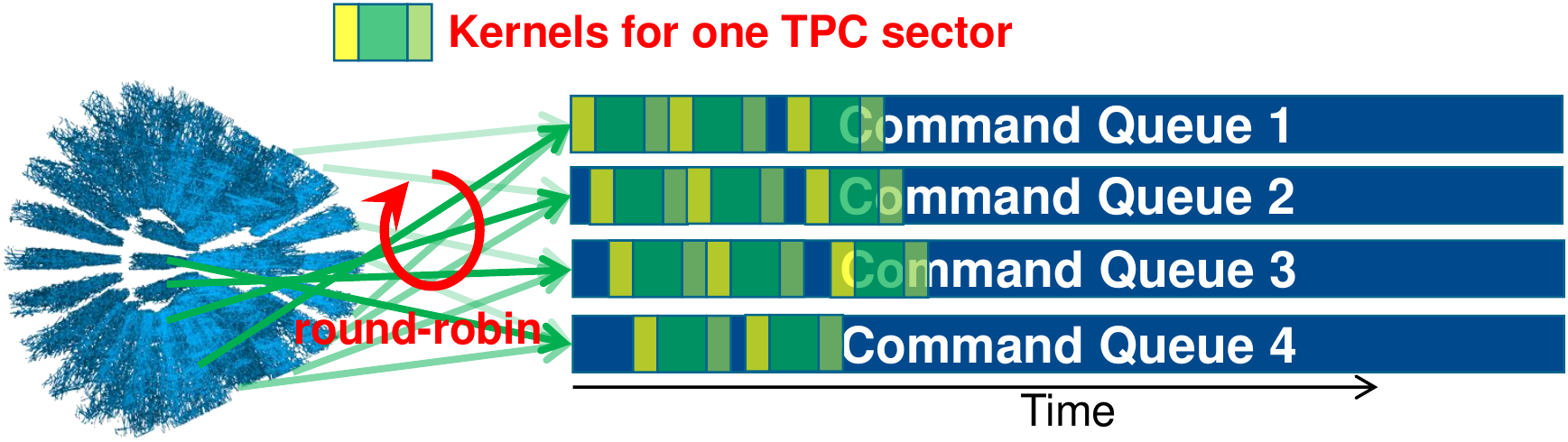}\end{center}
\begin{minipage}[b]{1.0\textwidth}\caption{Illustration of the alternate scheduling scheme for the GPU processing.
Multiple command queues are used in parallel.
The GPU scheduler creates parallelism by running multiple smaller kernels in parallel, compared to fewer large kernels in Figure~\ref{fig:async}.
Tasks for one TPC sector are sent to the same command queue, different TPC sectors are distributed into the queues in a round robin fashion.
The number of command queues is an optimization parameter, e.\;g.~eight queues is suited for the S9000 GPUs of the current HLT matching the number of the hardware queues in the GPU.}
\label{fig:a}
\end{minipage}
\end{figure}

The opportunity to have multiple kernels executed from multiple event queues in the same time enables additional parallelism inside an event.
The order of the kernel execution within a sector must be maintained, but different sectors are completely independent.
It is therefore possible to have multiple queues, and queue all kernels for one sector at once in one queue, but distribute the different sectors among multiple queues.
This works up to as many queues as there are sectors, although it is principally better to match the number of queues to the hardware constraints.
Figure~\ref{fig:a} illustrates the processing.

This approach leaves the parallelization to the GPU scheduler, and allows executing multiple of the cellular automaton kernels concurrently, or also parallelize the cellular automaton and the track following.
The data transfer from host to GPU is queued just before the kernels for a sector.
The transfer back to the host is a bit more complicated, because it is not known beforehand how many tracks are found.
There are two possible solutions: utilization of modern APIs where the GPU kernel itself can queue other GPU tasks: in this case a DMA transfer.
Or, the tracker can estimate a buffer that is large enough to store all tracks and queue the transfer of this buffer immediately after the last kernel.
In case the estimation was too small, the remaining tracks can be fetched afterwards.
We have not fully implemented this approach yet, but in a first stage we have implemented it for the first cellular automaton step which already reduces the required time on the HLT GPUs by approximately~$20$\,$\%$.

\subsection{Parallelize multiple events}

A totally different and simpler approach is the execution of kernels for multiple events concurrently.
The HLT GPU framework allows to start independent processing components, both performing track reconstruction on the same GPU, as long the GPU memory suffices for all of them.
This approach can also load the GPU well, but multiplies the memory requirement with the number of concurrent queues.

In the HLT, two instances of the GPU tracker running on the same GPU in parallel need in average~$220$\,ms to process a reference central heavy ion event, while a single exclusive instance finishes after~$145$\,ms (measured on the S9000 GPU~\cite{chep15}).
Hence, the parallel approach needs only~$110$\,ms per event reducing the overall processing time by~$24$\,$\%$.
The speedup is even larger for smaller events because they have even less parallelism.
This approach has been running in the HLT since the middle of~$2016$.
The aggregate reconstruction performance of the HLT cluster with~$180$ compute nodes enables the full reconstruction and fit of~$40.000.000$ tracks per second.

\section{Outlook for LHC Run~3}

\looseness=-1
Table~\ref{tab2} gives an overview over the maximum processing rate achieved in the HLT in data replay tests~\cite{framework}.
The second scenario processes pp events at maximum luminosity, the third one the largest Pb-Pb events with maximum pile-up.
Both run at the maximum rate the DDL\footnote{The DDL is the Detector Data Link that transfers incoming data in the HLT. This means the HLT is able to process data at the absolute maximum rate the experiment can deliver.} can deliver.
The fourth scenario uses the same events as the third one, but it skips all network transfer and all processing besides the TPC.
It runs only the TPC GPU tracking locally.
The comparison of the third to the forth shows the remaining GPU resources, which are still available under maximum load.
Leaving some margin, this means that the GPUs are used to roughly half capacity.

\begin{table}
\centering
\caption{\label{tab2}Maximum processing rate achieved in the HLT farm with data replay in several scenarios (see~\cite{framework}).}
\begin{tabular}{p{11.6cm}rl}
\br
Scenario & Rate & Limitation \\
\mr
pp (Pb-Pb Reference run, Run 244364, TPC, ITS, EMCAL, V0, ZDC) & $4.5$\,kHz & CPU \\
pp ($13$\,TeV, $25$\,ns, Run 239401, TPC, ITS, EMCAL, V0, ZDC) & $2.4$\,kHz & DDL$^1$ \\
Pb-Pb (Max. Luminosity, Run 245683, TPC, ITS, EMCAL, V0, ZDC) & $950$\,Hz & DDL \\
Pb-Pb (Run 245683, TPC only, no data transport) & $2.5$\,kHz & GPU \\
\br
\end{tabular}
\end{table}

The current HLT setup can fulfill all requirements for Run~2 and has currently around~$50$\,\% of GPU spare capacity available.
No additional improvement is required for Run~2, but the available resources can be used as test ground for new features needed for Run~3.
These new features can be tested already during Run~2 operation, and we will obviously use them already now if they improve the tracking results.

\subsection{Run~3 conditions}

For Run~3, the TPC tracking will face additional challenges:
\begin{enumerate}
 \item The TPC will be upgraded, Multi Wire Proportional Chambers (MWPC) will be replaced by Gas Electron Multipliers (GEM), and the readout will be continuous instead of triggered~\cite{gem}.
 \item The ALICE online system will not process individual events anymore but time-frames consisting of many interactions of accordingly larger data size.
 \item A time-frame contains information from many collisions. With respect to the TPC as a drift detector, the~$z$-coordinate of a hit can only be obtained from the drift time when the interaction time is known.
       It is thus not possible to transform all hits to spatial coordinates beforehand, but the vertex of the track needs to be identified first.
       Therefore, transformation and tracking can not run separatedly but are coupled.
\end{enumerate}

We do not see a significant problem in the larger data size of the time-frames compared to current event processing.
The size of a time-frame will be few gigabytes which already fits in the memory of state of the art GPUs, and this constraint will only relax with newer models and larger memory.
The additional memory the tracker needs for internal buffers is roughly three times the input data size.
If GPU memory would become insufficient, we can easily employ a slicing approach, processing parts of the time-frame one after another in a similar fashion as we currently process the TPC sectors.
Since the GPU tracking time goes linearly with the input data size, processing all events jointly in a time-frame should not take significantly longer than processing them individually.
To the contrary, the size of the time-frame offers much more parallelism than single events, which will help us use GPUs efficiently.
This is important, because it is unlikely that several time-frames will fit in GPU memory at the same time.

\begin{figure}[b]
\includegraphics[width=1.0\textwidth]{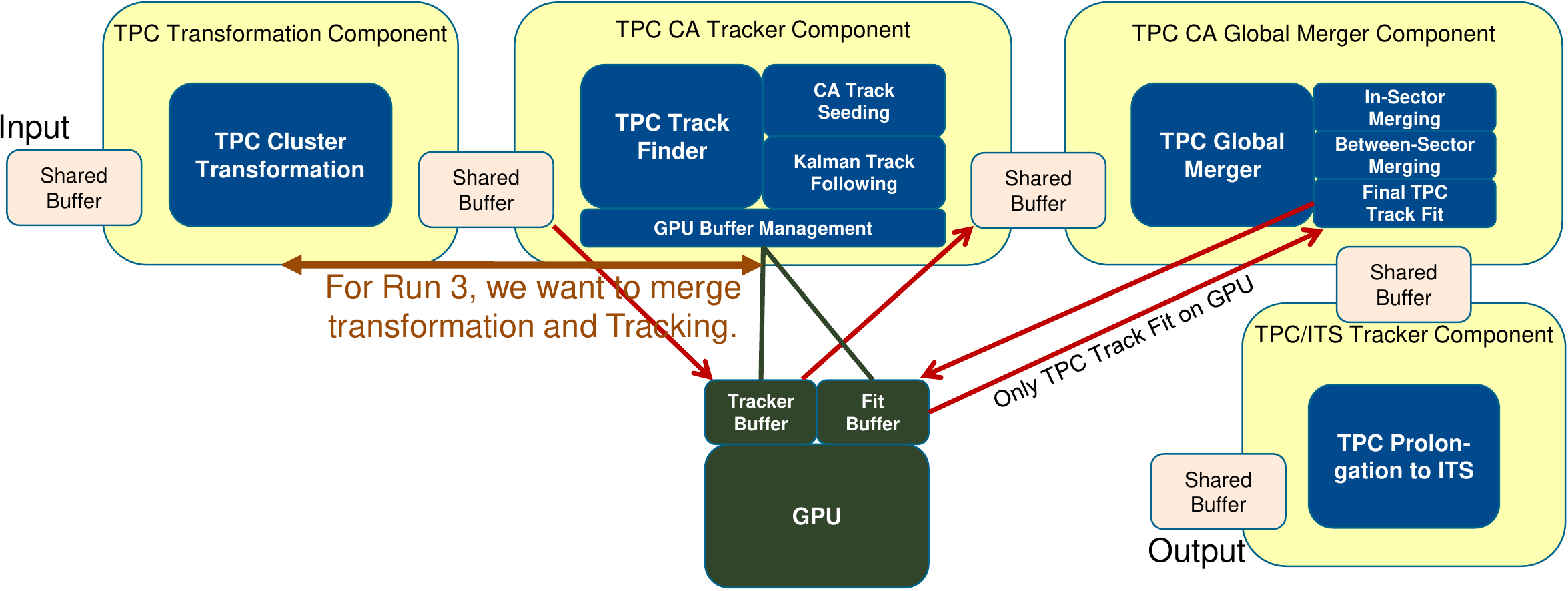}
\begin{minipage}[b]{1.0\textwidth}\caption{Illustration of all current HLT components related to TPC reconstruction.
The TPC cluster transformation applies the calibration and transforms the hits from TPC local coordinates into~$x$, $y$, $z$ coordinates for tracking.
The TPC CA Tracker Component and TPC CA Global Merger Component perform the two phases of the track reconstruction.
The TPC/ITS Tracker Component prolongs the TPC track into the ITS detector and improves the fit.
Components pass the data via shared memory buffers.
The TPC CA Tracker Component runs fully on the GPU.
The final track fit step of the TPC CA Global Merger Component can optionally run on the GPU, but this is currently not used due to the penalty of the additional data transfer.}
\label{fig:b}
\end{minipage}
\end{figure}

Currently, the transformation and tracking steps are completely isolated and run one after another.
Even further, the transformation runs on the host while the GPU performs the tracking.
This approach might be infeasible for Run~3 when they will be coupled, because it is prohibitively expensive to transfer data for the coordinate transformation during the tracking.
Figure~\ref{fig:b} visualizes the current situation.

\looseness=-1
The current HLT TPC tracking consists of four HLT components performing the transformation, the two phases of the actual TPC tracking, and finally the TPC / ITS propagation and refit.
Out of these four components, only one runs on the GPU: the track finding component (seeding and track following).
Since the track finding has by far the largest CPU load (see Table~\ref{tab1}), the GPU adaptation of other components will speed up the processing moderately, but it will free significant CPU resources for other tasks.
The track fit step inside the merger could optionally run on GPU, but this is currently inefficient due to the additional data transfer.
For efficient processing, this means that the first three components must be moved fully to the GPU to avoid any intermediate date transfer.
In this endeavor, it makes sense to move adjacent (data-dependency-wise) compute intense steps to the GPU as well, as the TPC / ITS tracking.

\begin{figure}[t]
\includegraphics[width=0.95\textwidth]{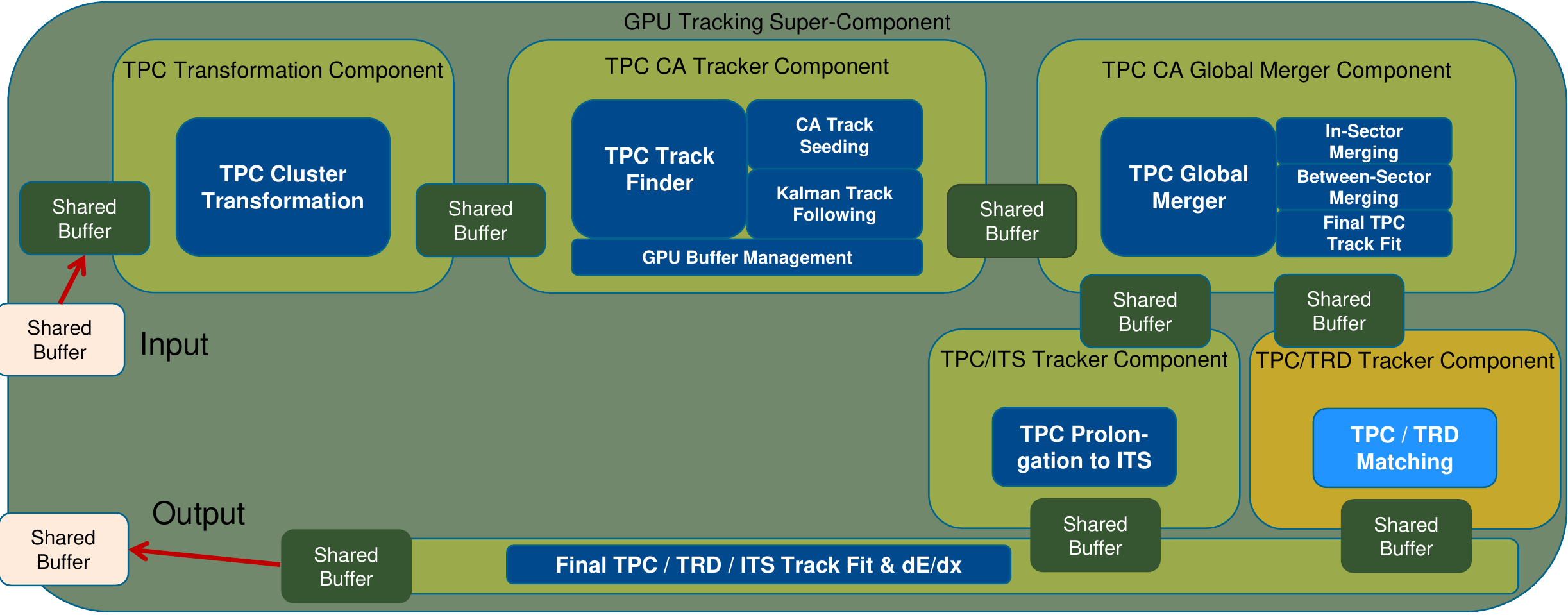}
\begin{minipage}[b]{1.0\textwidth}\caption{Illustration of a possible future TPC GPU reconstruction scheme.
More components are ported to the GPU, still communicating over shared buffers but inside GPU memory.
The component structure is maintained for better modularity, flexibility, and maintainability.
To the host, everything looks like a super-component performing all the steps at once.
Communication with other host components still happens via shared host buffers.
Data transfer to the GPU and from the GPU happens only once.}
\label{fig:c}
\end{minipage}
\end{figure}

Starting with the Run~2 software as basis, the first step is to move all components to the GPU, but leave the separation as is.
The integration of transformation and track finding will be handled in a later effort.
The HLT component structure has the advantage that all components can be tested individually, which simplifies debugging and improves maintainability.
Therefore, we do not want to change this approach.
We plan to replace the shared memory buffers on the host, which currently enable the data transfer between the components, by shared memory buffers on the GPU.
We will use a shared common source code supporting CPU and GPU as we do for the tracking already now~\cite{chep15}.
This gives us great flexibility.
For debugging purposes, we can still run any of the steps on the CPU, or compare CPU and GPU results of all steps individually.
For the final operation, everything will run on the GPU with only a single data transfer at the very beginning and the very end.
To the outside, this looks like a super-component that performs all steps at once.
Figure~\ref{fig:c} illustrates the new scheme.
Finally, we can add additional reconstruction components and features not yet available in the HLT in Run~2 like the TPC / TRD track prolongation and the dE/dx computation indicated in the figure.

\section{Conclusions}

We have shown that the ALICE HLT TPC GPU tracking is fast enough to cover all conceivable data taking scenarios for ALICE during HLT Run~2.
By improving the parallelization using new GPU features, we have already improved the tracking performance by~$24$\,\% and we expect a larger improvement for Run~3.
We have presented an approach to handle the TPC tracking in Run~3 on the GPU and how we want to extent the GPU reconstruction to other detectors and features.
The GPU tracking is running stable in 24/7 operation in the HLT.
For Run~2, no further improvements are necessary, but we will use the current HLT as test ground to commission new features for Run~3.

\end{document}